\documentclass[prb,twocolumn,superscriptaddress]{revtex4}
\usepackage{amsmath,amssymb,latexsym,epsfig,graphics,epsf}
\usepackage{graphics}
\usepackage{float}
\newcommand{\br}{\bold r}

\newcommand{\bR}{\bold R}

\newcommand{\tbR}{\tilde{\bold R}}

\newcommand{\tP}{\tilde{\Phi}}

\newcommand{\be}{\begin{equation}}
\newcommand{\ee}{\end{equation}}
\newcommand{\Scr}{S_{\rm cryst}}
\newcommand{\Sga}{S_{\rm gas}}
\newcommand{\Sgl}{S_{\rm glass, extr}}
\newcommand{\Sex}{S_{\rm ex}}
\newcommand{\Scon}{S_{\rm con}}

\begin{document}
\title{Entropy and relaxation time}
\author{Jeppe C. Dyre}\email{dyre@ruc.dk}
\affiliation{DNRF Center "Glass and Time", IMFUFA, Dept. of Sciences, Roskilde University, P. O. Box 260, DK-4000 Roskilde, Denmark}

\date{\today}

\begin{abstract}
This paper discusses the possible relation between entropy and the relaxation time of liquids, in particular glass-forming systems, providing supplementing comments to the paper entitled ``A brief critique of the Adam-Gibbs entropy model'' by Hecksher {\it et al.} [J. Non-Cryst. Solids {\bf 325}, 624-627 (2009)]. Besides the Adam-Gibbs model, we also briefly discuss Rosenfeld's excess entropy scaling and the 1964 experimental observation by Chang and Bestul of a universal excess entropy at the glass transition.
\end{abstract}
\maketitle

\section{Time and Entropy}\label{time}

Time and entropy are two of the most fundamental concepts of physics. Time and entropy are also key concepts of important unsettled scientific questions. Consider for instance the schism between reversibility and irreversibility: It is a law of nature that for any closed system entropy can only increase. In view of this, how can one understand the claim of modern physics that there is no fundamental irreversibility on the microscopic scale? In other words, why is it that entropy cannot decrease? I subscribe to the prevailing view that the irreversibility of macroscopic thermodynamics can always be traced back to unlikely initial conditions followed by a time evolution according to reversible laws, but famous scientists like Prigogine and Penrose do not think that things are this simple \cite{pri84,pen94}.

Another example where entropy meets time is in black-hole thermodynamics. A black hole sucks up anything that enters it, an irreversible process from which there is no escape. A great advance in the theory of black holes was the discovery that, nevertheless, a black hole has an entropy and an equilibrium temperature, and that thermal radiation can escape from the black hole's boundary, the so-called Hawking radiation \cite{haw75}.

Hawking radiation is a quantum phenomenon, and quantum mechanics is the final example that comes to mind when contemplating where in physics entropy meets relaxation time. There are just two fundamentally irreversible phenomena in physics: The law that entropy can only increase and the quantum-mechanical measuring process. The latter is the collapse of the wave function taking place during a measurement, after which the wave function corresponds to the eigenfunctions of the full set of eigenvalues being measured. There have been speculations of a connection between these two fundamental irreversibilities, but according to the conventional wisdom there is no such connection.

Given the above it is fascinating that our field of research provides further examples where entropy and time appear to be connected, and it is not surprising that such suggested connections for many years have generated great interest. This note discusses some of the suggested connections from a personal perspective, based on and expanding the paper \cite{dyr09} written jointly with Tina Hecksher and Kristine Niss for the proceedings of the First International Workshop on Glass and Entropy held in June 2008 in Slovakia. The present paper and the associated reprint focus on the Adam-Gibbs model \cite{ada65} and critiques of it. This model remains popular in the glass community, in a continuation of the efforts of Marty Goldstein long ago in understanding the glass transition in terms of entropy \cite{gol63}. Since then, notably, Austen Angell during almost 50 years \cite{ang68,ang74,ang95,ric98,ang00} -- like many of us -- was fascinated by the Adam-Gibbs model's intriguing suggestion of a connection between entropy and relaxation time and its prediction of an ideal glass state of zero configurational entropy and infinite relaxation time.

\section{Three suggested connections between entropy and relaxation time in liquids}\label{connec}

We use the term ``liquid'' for the equilibrated metastable thermodynamic state that has no memory of its thermal history and for which all properties are therefore uniquely defined functions of temperature and pressure; the term ``glass'' refers to the frozen out-of-equilibrium liquid phase. It is important to distinguish also between the different entropy concepts used: In standard liquid-state theory the term ``excess entropy'' $\Sex$ refers to the entropy minus that of the ideal gas at the same temperature and density, $\Sex(T,\rho)\equiv S(T,\rho)-\Sga(T,\rho)$ \cite{han13}. Since the latter quantity refers to the total chaos of molecular positions in a gas whereas any liquid has some degree of order, one always has $\Sex <0$. On the other hand, the tradition of glass science is to define the ``configurational'' entropy $\Scon$ as the actual entropy of the supercooled liquid minus the entropy of the crystalline state, $\Scon(T,p)\equiv S(T,p)-\Scr(T,p)$ in which $p$ is the pressure. Since the crystalline state is often not available, which is for instance the case for most polymers, one often calculates the configurational entropy by subtracting the glass entropy extrapolated to temperatures above the glass transition temperature $T_g$: $\Scon(T,p)\cong S(T,p)-\Sgl(T,p)$. The latter procedure is not unproblematic, however, because not only is the glass state not unique, the extrapolation itself can be done in different ways.

When discussing entropy in relation to a liquid's relaxation time as measured, e.g., via the diffusion constant or calculated as the inverse dielectric loss peak frequency, I can think of three important works. These are very briefly reviewed below. Throughout the discussion it is useful to keep in mind the distinction between ``ordinary'' liquids with relaxation times in the picosecond time range and ``viscous'' glass-forming, supercooled liquids for which the relaxation time spectrum extends over many decades, to hours, days, even weeks, depending on the patience of the experimenter. 

\subsection{Rosenfeld's excess entropy scaling}\label{ros}

Based on the  primitive computer simulations that were possible at the time, Rosenfeld in 1977 suggested that a liquid's reduced diffusion constant and reduced shear viscosity in the ``dense fluid region'' (i.e., non-gas-like states) is a universal function of the liquid's excess entropy \cite{ros77}. The term ``reduced'' refers to quantities measured in the unit system in which the length unit is $\rho^{-1/3}$, the energy unit is $k_BT$, and the time unit is $\rho^{-1/3}\sqrt{m/k_BT}$ where $m$ is the particle mass. It should be emphasized that these units are {\it not} those usually used for the Lennard-Jones (LJ) system -- the microscopic length $\sigma$ and the energy $\varepsilon$ of the LJ pair potential.

The systems studied by Rosenfeld were the LJ system, the one-component plasma, the $n=12$ inverse-power-law (IPL) pair potential, and the hard-sphere (HS) system. He justified the observed quasiuniversality of the excess (``internal'') entropy dependence of the reduced viscosity and diffusion constant by arguing that each of the systems studied is well represented by a HS system. Since the latter has just one degree of freedom as regards the configurational thermodynamics -- the packing ratio $\eta$ or, equivalently at a given density, the HS radius -- the excess entropy and transport coefficients are functions of $\eta$. Consequently, by elimination of $\eta$ the transport coefficients are functions of the excess entropy.

For many years there was only modest interest in Rosenfeld's observation of the importance of the excess entropy and the quasiuniversal way in which it appears to control liquid dynamics. During the last ten years, however, there has been a considerable focus on it, and 80\% of the paper's citations occurred the last ten years. Closely related to Rosenfeld's observation, but unaware of it, Dzugutov in 1996 noted that the two-particle entropy appears to control dynamic quantities like the self-diffusion constant \cite{dzu96}. In fact, the excess entropy consists of a sum of n-particle terms \cite{net58}, and the most important term is usually the two-particle entropy that is easily calculated from the radial distribution function.

During the last ten years there have been several investigations of Rosenfeld excess entropy scaling, mostly based on computer simulations. These studies have generally confirmed its usefulness, compare Refs. \onlinecite{err06,mit06a,ruc06,aga11} that give references to other relevant papers.

\subsection{The Chang-Bestul connection}

In a paper from 1964 entitled ``Excess entropy at glass transformation'' Chang and Bestul compiled results from careful measurements on a number of glass-forming liquids \cite{bes64}. These authors define the excess entropy as the entropy difference between the supercooled liquid at a given temperature and that of the ``supercooled amorphous phase in a unique reference state from which, even in principle, no further entropy can be lost by liquid-like configurational relaxation''. Today one would refer to this as the ideal glass state of zero configurational entropy, by some believed to be approached for the equilibrium liquid if the temperature approaches the VFT divergence temperature $T_0$. Note that this definition is different from the one usually used in liquid state theory.

The conclusion based on the data compiled by Chang and Bestul was that the excess entropy ``per bead'' (equivalent to polymer molecular chain lengths as defined by Wunderlich \cite{wun60}) is an almost universal constant at the calorimetric glass transition.

There is nothing special about the calorimetric glass transition, which usually corresponds to cooling rates of order Kelvin per minute -- for a different cooling rate the glass transition takes place at a (slightly) different temperature. Consequently, the reported universality of the excess entropy at $T_g$ should apply also for other cooling rates. The glass transition takes place when the equilibrium liquid's relaxation time $\tau$ upon cooling becomes comparable to the inverse cooling rate $\tau\sim 1/(d\ln T/dt)$. The logical consequence of the Chang-Bestul work is therefore that the liquid's relaxation time is a universal function of the excess entropy.

\subsection{The Adam-Gibbs model}

The most popular entropy model is that of Adam and Gibbs from 1965 \cite{ada65}, which is the subject of our 2009 paper ``A brief critique of the Adam-Gibbs entropy model'' \cite{dyr09}. The Adam-Gibbs idea is that the loss of configurational entropy upon cooling -- in practice identified with the excess entropy as Bestul and Chang defined it experimentally -- means that the system has much fewer states to jump between as the temperature is decreased. A state is here to be thought of as a potential-energy minimum, an inherent state, although the Adam-Gibbs paper preceded the  introductions and discussions of these concepts by Goldstein in 1969 \cite{gol69} and by Stillinger and Weber in 1983 \cite{sti83}. The loss of available states upon cooling according to Adam and Gibbs means that larger and larger ``cooperatively rearranging regions'' must be involved in transitions between any two states. Assuming that the activation energy is proportional to the volume of these regions, Adam and Gibbs arrived at

\be\label{ada}
\tau(T)=\tau_0\exp\left(\frac{A}{TS_c(T)}\right)\,.
\ee
In particular, the relaxation time diverges at the Kauzmann temperature, the temperature at which the excess entropy extrapolates to zero. The beautiful physical picture is that at this temperature there is just one equilibrium liquid state and for this reason, no transitions are possible. In this way the Adam-Gibbs picture was very modern in ascribing the dramatic increase of the relaxation time observed upon cooling for a glass-forming liquid to the approach to a second-order phase transition. Thus there is an underlying phase transition that can never be reached, the existence of which nevertheless is the cause of the dramatic slowing down. This idea remains popular in the glass community, also among leading theorists as exemplified, e.g., in the random first order transition (RFOT) theory of Wolynes and coworkers \cite{lub07,wol12}.

\section{Critiques of the Adam-Gibbs entropy  model}\label{ag_en}

Reference \onlinecite{dyr09} from 2009 summarizes our pragmatic critiques of the Adam-Gibbs model, the details of which will not be repeated here. The critiques are grouped into two categories. Regarding the {\it model assumption} there are a number of {\it ad hoc} assumptions one may question and wonder about: 

\begin{itemize}
\item That the liquid may be regarded as a collection of independently rearranging regions; 
\item That a region must have at least two configurations;
\item That the activation energy is proportional to the region volume;
\item That a unique ``ideal'' glass state exists even though no characterization of it has been given. 
\end{itemize}

Regarding the Adam-Gibbs model's {\it experimental validation} one may question and wonder about: 
\begin{itemize}
\item The small sizes of the assumed independently rearranging regions arrived at in fitting model to data (typically 4-8 molecules at the lowest temperatures, smaller at higher temperatures);
\item The identification of the configurational entropy with the experimental excess entropy, a quantity that is, moreover, not even experimentally well defined;
\item The assumption that the crystal and the glass(es) have identical vibrational entropies often used to calculate the configurational entropy;
\item The assumption that a liquid cannot have a lower entropy than the crystal at the same temperature;
\item The ``Kauzmann'' extrapolation of the supercooled liquid's entropy to temperatures below $T_g$;
\item The basis for identifying the Kauzmann temperature at which the extrapolated excess entropy vanishes with the VFT temperature at which the liquid's relaxation time diverges \cite{tan03}.
\item The basis for concluding that the equilibrium liquid's relaxation time diverges at a finite temperature \cite{hec08};
\end{itemize}

Despite the above listed potential problems, the Adam-Gibbs model remains popular for the simple reason that many data have been reported to conform to Eq. (\ref{ada}). These include extensive experimental data on organic \cite{ric98} and inorganic glass formers \cite{ric95}, as well as extensive computer simulation data \cite{sas00,kar09,sta13}. 

We proceed to review a recent theoretical development that throws light on the role of entropy for the dynamics of liquids.

\section{Insights from the isomorph theory}\label{isom_les}

We start by briefly reviewing the isomorph theory that was proposed in 2009 in Ref. \onlinecite{IV}; the theory was summarized in more detail with a focus on its experimental consequences in Ref. \onlinecite{ped11} as well as with a more theoretical focus in Ref. \onlinecite{ing12}.
 
An isomorph is a curve in the two-dimensional thermodynamic phase diagram -- parameterized, e.g., by density and temperature or by pressure and temperature -- along which several quantities are invariant when given in the reduced units defined in Sec. \ref{ros}. The isomorph invariants include the excess entropy, the isochoric specific heat, the instantaneous shear modulus, the diffusion constant, the viscosity, in fact structure and dynamics as such are invariant in reduced units. Isomorphs are configurational adiabats, but it is not the case that configurational adiabats for all systems have the several additional invariants that characterize isomorphs. In other words, an isomorph is a configurational adiabat, but not all such adiabats are isomorphs.

Consider a classical-mechanical system of $N$ particles of mass $m$ in volume $V$ with density $\rho=N/V$. The collective position vector is defined by $\bR\equiv (\br_1,...,\br_N)$ and the reduced collective position vector is defined by $\tbR\equiv \rho^{1/3}\bR$. Two state points with density and temperature $(\rho_1,T_1)$ and $(\rho_2,T_2)$ are isomorphic \cite{IV} if whenever two of their microconfigurations have the same reduced coordinates, their canonical probabilities are the same. This translates into $\exp\left(-{U(\rho_1^{-1/3}\tbR)}/{k_BT_1}\right)=C_{12}\exp\left(-{U(\rho_2^{-1/3}\tbR)}/{k_BT_2}\right)$ in which $\tbR$ is the common reduced position vector of the two microconfigurations. The equivalence classes of this relation are the system's isomorphs \cite{IV}. The only systems that have exact isomorphs are those for which the potential energy is a homogeneous function,  i.e., obeys $U(\lambda\bR)=\lambda^{-n}U(\bR)$ for some $n$, for instance systems with inverse-power-law (IPL) pair potentials.

A system has isomorphs if and only if in the relevant part of thermodynamic phase space the system has strong correlations between its $NVT$ (canonical) equilibrium fluctuations of virial and potential energy \cite{IV}, typically with a Pearson correlation coefficient above  $0.9$. This is never the case close to (most of) the gas-liquid coexistence line, at the critical point, or in the gas phase; strong virial potential-energy correlations are found at typical condensed-phase liquid state points -- as well as at all crystalline and glass states \cite{II,cryst}. 

The class of liquids with good isomorphs in the condensed-phase part of phase space was previously referred to as ``strongly correlating'', but since this was repeatedly confused with strongly correlated quantum systems and since these liquids are simple in their properties \cite{ing12}, we now call them ``Roskilde-simple''.

Computer simulations have demonstrated several examples of Roskilde-simple systems, for instance \cite{I,ing12}: The Lennard-Jones (LJ) system and its generalizations to mixtures and to other exponents than 6 and 12, simple molecular liquids like the asymmetric dumbbell or the Wahnstrom OTP model, the Buckingham liquid, the ``repulsive'' LJ system (i.e., with plus instead of minus between the two IPL terms), the 10-bead flexible LJ chain \cite{vel13}, etc. The isomorph theory  works very well for solids; thus an LJ crystal has more than 99\% $WU$ correlations and, e.g., the phonon spectrum and the vacancy diffusion constant are isomorph invariants \cite{II,cryst}. 

For real systems, we believe that most or all van der Waals bonded and metallic liquids are Roskilde-simple \cite{II}, whereas covalently- and hydrogen-bonded liquids are not because strong directional bonding usually ruin strong virial potential-energy correlations and thus the isomorphs. 

The isomorph theory explains a number of experimental observations, for instance density scaling of the dynamics for van der Waals liquids and why this works poorly for hydrogen-bonded liquids \cite{rol05}, or isochronal superposition for van der Waals liquids, the fact that the relaxation time for some liquids determines the entire relaxation spectrum throughout the thermodynamic phase diagram \cite{nga05}. The melting line is an isomorph, and this explains the invariances along this line of reduced viscosity, excess entropy, etc \cite{IV,V}. A further confirmation of the isomorph theory was recently provided by the experimental validation of the theory's prediction that the density-scaling exponent can be determined from linear-response thermoviscoelastic measurements at ambient pressure \cite{gun11}.

Roskilde-simple systems have simple thermodynamics. Thus if $s$ is the excess entropy per particle, temperature factorizes as follows \cite{ing12a} $k_BT=f(s/k_B)h(\rho)$ in which $f$ is a dimensionless function. Since the excess entropy is isomorph invariant, the isomorphs are given by ${h(\rho)}/{T}={\rm Const}$. For systems described by pair potentials of the form $v(r)=\sum_n \varepsilon_n (r/\sigma)^{-n}$ the function $h(\rho)$ is given by $h(\rho)=\sum_n C_n \rho^{n/3}$ in which the only non-zero terms are those that occur in $v(r)$ \cite{ing12a,boh12}. In conjunction with ${h(\rho)}/{T}={\rm Const}$ this provides a convenient way to map out the isomorphs in the phase diagram.

It was recently argued \cite{dyr13a} that a Roskilde-simple system is characterized by ``hidden scale invariance'' manifested in the approximate relation 

\be
U(\bR)\cong h(\rho)\tP(\tbR)+g(\rho)\,.
\ee
Clearly, all the reduced-unit physics is contained in the term $\tP(\tbR)$; since it is dimensionless, this term cannot involve the characteristic length $\sigma$ of the microscopic potential or the microscopic energy scale $\varepsilon$. The overall energy scale is set by $h(\rho)$ -- even  away from thermal equilibrium -- which for some dimensionless function $\tilde h$ can be written $h(\rho)=\varepsilon\tilde h(\rho\sigma^3)$.

Returning now to the connection between entropy and relaxation time of liquids, we first note that the isomorph theory gives rise to an ``isomorph filter'' \cite{IV}, which can be used to distinguish between different theories of a liquid's 
relaxation time. Suppose a theory for glass-forming liquids' relaxation time claims universal validity, like the Adam-Gibbs model \cite{ada65}, the free-volume theories, the shoving model \cite{dyr96,dyr06}, etc. Then the theory must also apply for the Roskilde-simple liquids. Since the difference between reduced and non-reduced relaxation time is insignificant for a supercooled liquid, and since the (reduced) relaxation time is an isomorph invariant, the quantity controlling the relaxation time must also be an isomorph invariant. This is the case for Rosenfeld's excess entropy scaling which states that the relaxation time is function of the excess entropy. Likewise, this is the case for the shoving model because the reduced instantaneous shear modulus is an isomorph invariant \cite{IV}. If the number $A$ of the Adam-Gibbs relation Eq. (\ref{ada}) is constant throughout the phase diagram, however, as is usually assumed in experimental validations of this relation, the Adam-Gibbs relation does not survive the isomorph filter, i.e., it cannot apply universally. The computer simulations confirming this relation do reveal a density dependence of $A$, and interestingly it has very recently been verified that this density dependence is precisely large enough to make the Adam-Gibbs relation consistent with the isomorph filter ($A\propto\rho^\gamma$ in which $\gamma$ is the density-scaling exponent) \cite{sen13}. 

The isomorph theory puts the entropy-time discussion into a new perspective. About half of all liquids (metals and van der Waals systems) are Roskilde-simple in the condensed phase including close to the glass transition, and for any such system the relaxation time is a unique and well-defined function of the excess entropy. This does not imply a causal relation between relaxation time and entropy, however; any other isomorph invariant also determines the relaxation time, for instance the isochoric specific heat, as recently demonstrated \cite{ing13a}. 

The theory of Roskilde-simple liquids' isomorphs reminds us that a {\it correlation} between two quantities does not logically imply a {\it causal relationship}. In this respect, I always come to think of the fact that most people die in a bed. This does not imply that it is dangerous to go to bed, of course; on the contrary it may be dangerous not to do so for a seriously ill person!

\section{Where does all this leave us?}\label{disc}

{\it Does entropy control the relaxation time of a glass-forming liquid?} Consider first the fairly large class of Roskilde-simple liquids. Here one cannot give a definite answer to the question, because any isomorph invariant will appear to ``control'' any other. Thus the excess entropy appears to control the relaxation time for these liquids, but so does, e.g., the isochoric specific heat or the reduced instantaneous shear modulus. In fact, one might equally well state that the relaxation time  ``controls'' the excess entropy! The situation is similar to that of the Lindemann melting criterion for a Roskilde-simple crystal: The crystal's reduced vibrational mean-square displacement is an isomorph invariant, i.e., invariant along the melting line. For this reason it will appear as if the crystal melts {\it because} the vibrations reach a certain strength, which makes a lot of sense physically. But according to the isomorph theory there need not be a compelling causal connection.

Consequently, in order to answer the above question one must focus on non-Roskilde-simple glass formers, i.e., on hydrogen-bonded and covalently bonded liquids. It is noteworthy that the Adam-Gibbs expression has been reported to fit data also for such systems, a fact that certainly strengthens the connection between entropy and relaxation time. On the other hand, the situation is not quite clear to me since, for instance, Rosenfeld excess entropy scaling appears to work best for Roskilde-simple liquids. More work is needed to resolve this issue, in truth a very common conclusion to a  scientific enquiry. -- Discussions on the intriguing connection between entropy and time in supercooled liquids will most likely continue for a number of years.

\acknowledgments 
This paper was presented at the {\it Symposium on Fragility} held January 5-8, 2014, Bangalore, as a tribute to glass science's grand old man Austen Angell. Austen is a polyhistor of the natural sciences, and his eternal curiosity and search for unexpected links between different fields and observations remain a strong source of inspiration to all of us. -- The centre for viscous liquid dynamics ``Glass and Time'' is sponsored by the Danish National Research Foundation via grant DNRF61.

\end{document}